\documentclass[iop]{emulateapj}
\usepackage{natbib}
\usepackage{mathptmx} 
\usepackage[T1]{fontenc}
\usepackage{graphicx}
\usepackage[colorlinks=true,urlcolor=blue,citecolor=blue,linkcolor=blue]{hyperref}
\def\kms {{\mathrm{km}\,\mathrm{s}^{-1}}}
\def\caii {Ca\,\textsc{ii}}
\defcitealias{Pereira:2012spic}{\mbox{Paper I}}

\shorttitle{The Effects Of Resolution On Spicule Properties}
\shortauthors{Pereira, De Pontieu, \& Carlsson}

\begin{document}

\title{The Effects of Spatio-temporal Resolution on Deduced Spicule Properties}
\author{Tiago M. D. Pereira$^{1,2}$, Bart De Pontieu$^{2}$, Mats Carlsson$^{3,4}$}
\affil{$^{1}$NASA Ames Research Center, Mof{}fett Field, CA 94035, USA;\\
$^{2}$Lockheed Martin Solar and Astrophysics Laboratory, 3251 Hanover Street, Org. A021S, Bldg. 252, Palo Alto, CA 94304, USA\\
$^{3}$Institute of Theoretical Astrophysics, P.O. Box 1029, Blindern, N-0315 Oslo, Norway\\
$^{4}$Center of Mathematics for Applications, University of Oslo, PO Box 1053, Blindern, 0316 Oslo, Norway
}

\begin{abstract}
Spicules have been observed on the sun for more than a century, typically in chromospheric lines such as H$\alpha$ and \caii\ H. Recent work has shown that so-called `type II' spicules may have a role in providing mass to the corona and the solar wind. In chromospheric filtergrams these spicules are not seen to fall back down, and they are shorter-lived and more dynamic than the spicules that have been classically reported in ground-based observations. Observations of type II spicules with \emph{Hinode} show fundamentally different properties from what was previously measured. In earlier work we showed that these dynamic type II spicules are the most common type, a view that was not properly identified by early observations.
The aim of this work is to investigate the effects of spatio-temporal resolution in the classical spicule measurements.
Making use of \emph{Hinode} data degraded to match the observing conditions of older ground-based studies, we measure the properties of spicules with a semi-automated algorithm. These results are then compared to measurements using the original \emph{Hinode} data.
We find that degrading the data has a significant effect on the measured properties of spicules. Most importantly, the results from the degraded data agree well with older studies (\emph{e.g.} mean spicule duration more than 5 minutes, and upward apparent velocities of about $25\;\kms$).
These results illustrate how the combination of spicule superposition, low spatial resolution and cadence affect the measured properties of spicules, and that previous measurements can be misleading.

\end{abstract}


\section{Introduction}

Spicules are dynamic structures in the solar chromosphere, first observed by \citet{secchi1877}. Traditionally observed in H$\alpha$, many of the earlier studies were reviewed by \citet{Beckers:1968,Beckers:1972}. These reviews painted a classical view of spicules as jets ascending with apparent velocities around $25\;\kms$, lifetimes of about 5~min, and reaching heights of $6-9\;$Mm. For several decades, this observational description has provided the constraints for spicule formation mechanisms \citep{Sterling:2000}. The reported lifetimes of 5~min suggested a connection to the photospheric/granular time scales. The coronal heights reached made spicules candidates to transfer mass and energy from the photosphere to the corona \citep{Beckers:1968,PneumanKopp:1978,Athay:1982,Tsiropoula:2004}. Yet these enigmatic structures have eluded modelers partly because, as noted by \citet{Sterling:2000}, older observations were not detailed enough and spicules were just beyond the resolution limits of instruments. 

The launch of \emph{Hinode} \citep{Kosugi:2007} and the combined use of adaptive optics and image post-processing \citep{vanNoort:2005} in ground-based observations brought a breath of fresh air to the observational studies of spicules. By providing a seeing-free, stable observing platform with high spatial and temporal resolution, \emph{Hinode} showed spicules in unprecedented detail. Using \emph{Hinode} observations, \citet{DePontieu:2007} reported spicules as much more dynamic features than previously thought, and suggested the existence of two types of spicules: slower, longer-lived `type I' and faster, shorter-lived `type II'. Recently, \citet[hereafter \citetalias{Pereira:2012spic}]{Pereira:2012spic} have extended this work by measuring a comprehensive set of spicules over several solar regions observed with \emph{Hinode}. \citetalias{Pereira:2012spic} confirmed the existence of two types of spicules, finding type I spicules to have typical lifetimes of $150-400\;$s, ascending velocities of $15-40\;\kms$, and type II spicules to have lifetimes of $50-150\;$s, and velocities of $30-130\;\kms$. Besides the lifetimes and velocities, the major difference between type I and type II spicules is that the former rise and fall in a parabolic trajectory, while the latter only have a rise phase and then fade from the passband.

The fading of type II spicules has fueled the idea that they are rapidly heated to higher temperatures not visible in the \caii\ H filter, and may deposit mass and energy at coronal heights \citep{DePontieu:2007,DePontieu:2009}. Given the mass flux transported by spicules, even if only a few percent of the mass flux reaches coronal temperatures, spicules can help power the solar wind and energize the corona \citep{Beckers:1968,PneumanKopp:1978,DePontieu:2009}. This view of type II spicules being heated to coronal temperatures has been reinforced by \citet{DePontieu:2011}, who find that some type II spicules, after fading from the \caii\ H passband, appear in transition region and coronal temperature filtergrams from the Atmospheric Imaging Array (AIA) on board the Solar Dynamics Observatory (SDO). Thus, it is important to make the distinction between type I spicules (whose material falls back after it rises) and type II spicules, which have more potential to mediate the transfer of energy and mass from the chromosphere to the corona.

While type I spicules have properties consistent with the `classical' spicules of \citet{Beckers:1968,Beckers:1972}, type II spicules do not. Therefore, the question arises: why are measurements of classical spicules seemingly incompatible with type II spicules seen from \emph{Hinode}? The question is more pertinent as both \citet{DePontieu:2007} and \citetalias{Pereira:2012spic} find type I spicules abundant only in active regions, with \citetalias{Pereira:2012spic} finding type II spicules dominant in the quiet sun and coronal holes. A plausible explanation is that the lower spatio-temporal resolution used before precluded the detection of the highly dynamic type II spicules. With such conditions, it is possible that spicules appearing and disappearing at the same place might be interpreted as one longer lived spicule, thereby biasing the measurements towards longer lifetimes and presumably different velocities. In this work, we put this hypothesis to the test by using the same spicule detection methods of \citetalias{Pereira:2012spic} on the same \emph{Hinode} data but now degraded to approximate the conditions of previous studies. We then compare the results of spicules measured on the degraded data with those from \citetalias{Pereira:2012spic}. Our aim is to study the effects of the spatio-temporal resolution in spicule measurements, in particular those of early observations that provided the basis for the `classical' view on spicules.

\section{Analysis and Results}

The data used are the quiet sun subset of \citetalias{Pereira:2012spic}. They span four sets of observations with \emph{Hinode}/SOT Broadband Filter Imager \citep{Tsuneta:2008,Suematsu:2008}, in the \caii\ H (396.85~nm) filter. These sets were chosen for their high cadence and quietness (\emph{e.g.}, absence of brightenings in X-ray images). The data were reduced using the standard IDL procedure \texttt{fg\_prep.pro}, and co-aligned using the cross-correlation technique of \citet{DePontieu:2007}. A radial density filter \citep[as in][]{DePontieu:2007,Okamoto:2007} was applied, along with a diagonal difference (emboss) filter to enhance the visibility of the spicules. Details of the observations are listed in Table~\ref{tab:obs}.

\begin{deluxetable}{lrrrrrr}
\tablecaption{Observational data sets.\label{tab:obs}}
\tablehead{
\colhead{Starting time} & \colhead{x coord.} & \colhead{y coord.} & \colhead{$\Delta t$} & \colhead{Duration} & \colhead{$N_{\mathrm{O}}$} & \colhead{$N_{\mathrm{D}}$} \\
     & \colhead{(arcsec)} & \colhead{(arcsec)} & \colhead{(s)} & \colhead{(min)} & &}
\startdata
2007-04-01T14:09 & $ 900$ & $-288$ & $4.8$ & $55$  & $24$ &  $8$ \\
2007-04-01T15:09 & $ 933$ & $-145$ & $4.8$ & $55$  & $47$ &  $7$ \\
2007-08-21T13:35 & $ 945$ & $   0$ & $4.8$ & $204$ & $62$ & $25$ \\
2011-09-16T03:00 & $-955$ & $ 259$ & $3.2$ & $66 $ & $41$ & $12$ 
\enddata
\tablecomments{$\Delta t$ is the observational cadence, $N_{\mathrm{O}}$ the number of spicules detected in the original data, and $N_{\mathrm{D}}$ the number of spicules detected in the degraded data.}
\end{deluxetable}

\begin{figure}
\begin{center}
\includegraphics[scale=0.95]{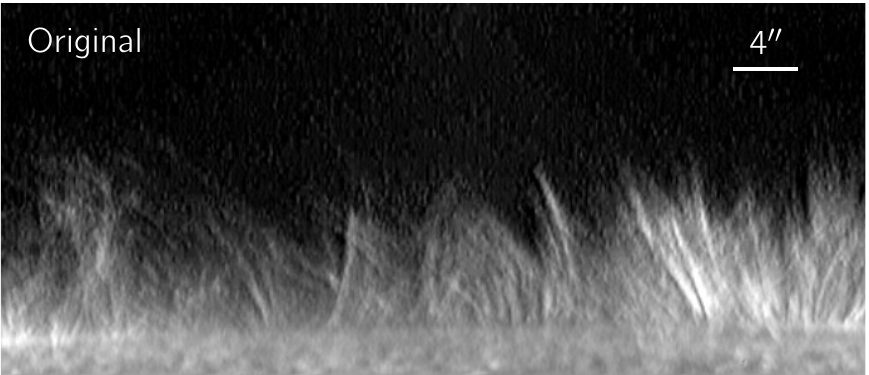}\\
\includegraphics[scale=0.95]{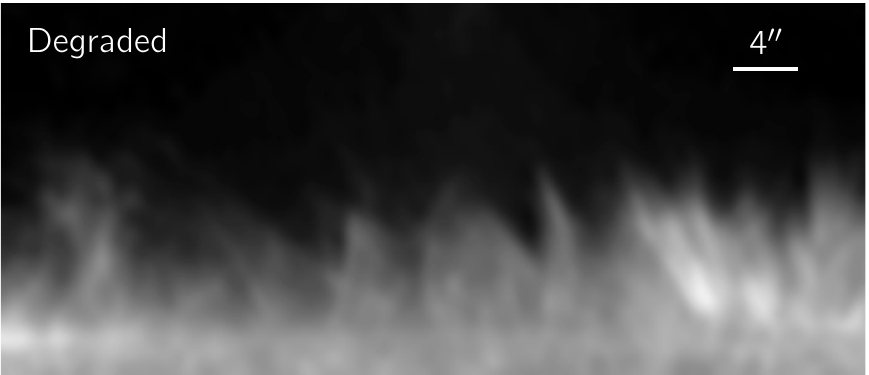}
\end{center}
\caption{Sample images from 2007 August 21, 14:33:39 UT. The degraded image (bottom) was obtained by blurring the original image (top) with a Gaussian filter with a FWHM of $1\arcsec$. Radial density and emboss filters were applied to the images. Each panel is approximately $56\arcsec\times24\arcsec$, and the approximate solar (x, y) coordinates for the lower left corner are ($935\arcsec.9$, $28\arcsec.1$). [\emph{This figure is available as an animation in the electronic edition of the Journal.}]\label{fig:spicules}}
\end{figure}

To compare spicule measurements with previous studies we have degraded the \emph{Hinode} data to mimic older observations. The spatial resolution has been degraded with a $1\arcsec$ FWHM Gaussian, and we sampled the images in time to get a cadence of $38.4\;$s. The same radial density and emboss filters were applied to the degraded data. The effect of the spatial degradation is illustrated in Fig.~\ref{fig:spicules}. Meant to encompass the heterogeneous range of parameters of older observations, these values are a simple approximation. Some studies had higher spatial resolution, others higher cadence (see Table~\ref{tab:spic_others}), but overall we believe our values to be a reasonable approximation, in particular because other effects such as seeing were not introduced in the degradation. In this kind of filtergram analysis, seeing can have two main effects: a reduced spatial resolution from a consistent less than ideal seeing, and variable time series quality when the seeing varies significantly with time. With our data degradation, the spatial resolution effects of a constant seeing are accounted for, but not the time series effects. One should note that the effects of seeing on time series are probably often underestimated: only a few bad frames in a sequence with good seeing can be very disruptive for dynamic events such as spicules.

For our comparison we use the quiet sun results of \citetalias{Pereira:2012spic} referring to them as `original', along with the results from the `degraded' data. For consistency, we used the same semi-automated algorithm, procedure, and settings of \citetalias{Pereira:2012spic} for detecting and measuring spicules. The method is briefly outlined below, with a more detailed description in \citetalias{Pereira:2012spic}.

The spicule detection is based on the algorithm used by \citet{vdVoort:2009} and tracks the time evolution of spicules, detecting spicules in each image and linking them with matching features in following images. From the detection algorithm we obtain the coordinates of spicules linked across several images. These chains of events are visually inspected and purged of false positives (\emph{e.g.} detections that are not spicules or spicules that are incorrectly linked). The next step is to measure the properties of spicules.

\begin{figure*}
\begin{center}
\includegraphics[scale=0.95]{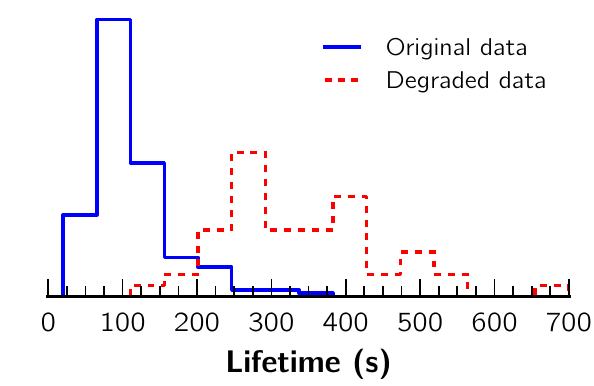}\includegraphics[scale=0.95]{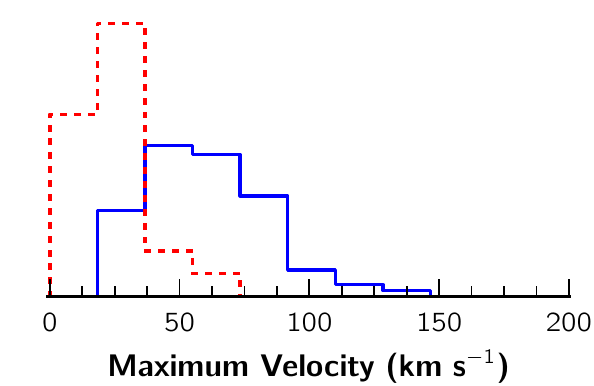}\includegraphics[scale=0.95]{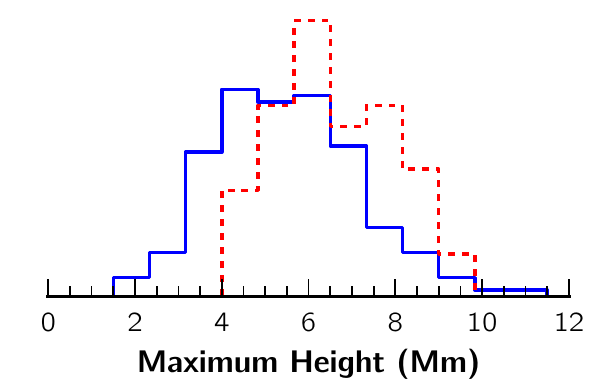}
\end{center}
\caption{Normalized distributions of spicule properties for the original (blue solid) and degraded (red dashed) data.\label{fig:hists}}
\end{figure*}

From the several quantities measured in \citetalias{Pereira:2012spic}, here we focus only on three: lifetime, maximum ascending velocity, and maximum height. The lifetime is obtained by visual inspection of the images before and after those in which a spicule was detected. The frames where the spicule is first and last seen are identified, and the lifetime is given by the time interval between them. Similarly, the maximum height is obtained by visual inspection of the images where the spicule is seen. The image were the spicule reaches its maximum height is identified and the height is measured as the closest distance from the top of the spicule to the limb. Finally, the maximum velocity is derived using space-time diagrams of the spicule intensity along its axis. Using the spicule coordinates provided by the algorithm, space-time diagrams are built accounting for the spicular transverse motion. The approximate positions of the spicule tops for each time are identified in this diagram, and the velocity is extracted from a linear fit to these points.

Measuring limb spicules is a difficult task, which leads to a certain subjectivity in the measurements when the signal is not very clear. The large degree of spicule superposition at the limb is the main observational difficulty in tracking and measuring these objects. As discussed in \citetalias{Pereira:2012spic}, our spicule detection method has uncertainties and selection effects, but is nevertheless a consistent way to compare spicules across data sets and better than a purely visual estimation, in particular for comparing spicule properties in the original and degraded data. 

In the degraded data we detected a total of 52 spicules. Of these, we derived the velocities from only 23 spicules, which provided the clearest signal. This is perhaps too conservative, but we wanted to keep our sample as clear as possible from less accurate measurements. From Table~\ref{tab:obs} one can see that in the original data over three times more spicules were detected. This discrepancy comes largely from the spatial degradation, whose blurring lowers the contrast of spicules and makes the automated detection harder. In \citetalias{Pereira:2012spic} more than 98\% of the spicules in these observations were found to be of type II, and the remaining of type I. In the smaller numbers of spicules detected in the degraded data, we did not find spicules with a parabolic rise and fall motion (as type I spicules), or spicules with a descending phase.

The distributions of the measured spicule properties are shown in Fig.~\ref{fig:hists}. For the degraded data we obtain a mean lifetime of $340\;$s with a standard deviation of $\sigma=107\;$s; a mean maximum velocity of $25.8\;\kms$ with $\sigma=11.4\;\kms$; and a mean maximum height of $6.66\;$Mm with $\sigma=1.35\;$Mm. For the original data the results were a mean lifetime of $111\;$s with $\sigma=54\;$s; a mean maximum velocity of $60.0\;\kms$ with $\sigma=23.3\;\kms$; and a mean maximum height of $5.51\;$Mm with $\sigma=1.70\;$Mm.

\begin{deluxetable}{lccccc}
\tablecaption{Comparison of mean spicule properties.\label{tab:spic_others}}
\tablehead{
Observations & \colhead{Cadence} & \colhead{$N$} & \colhead{$\langle t\rangle$} & \colhead{$\langle v_{\mathrm{up}}\rangle$} & \colhead{$\langle H_{\mathrm{max}}\rangle$} \\
                      &  \colhead{(s)}    &               &  \colhead{(s)}     &        \colhead{($\kms$)} &             \colhead{(Mm)}}
\startdata 
\citet{Roberts:1945}  &             $60$  &         $48$  &            $270$  &                      $30$ &                  $7.3$  \\
\citet{Dizer:1952}    &             $30$  &        $353$  &            $132$  &                   $20,\;40$\tablenotemark{a} &    $7.4$ \\
\citet{Rush:1954}     &          $10-60$  &        $101$  &            $210$  &                      $31$ &                     $9.0$ \\
\citet{Lippincott:1957}&    $\approx 30$  &         $77$  &            $342$  &                      $24$ & $7.3,\;9.5$\tablenotemark{b} \\
\citet{Nishikawa:1988} &    $\approx 30$  &          $4$  &            $339$  &                      $42$ &                     $8.7$ \\
\citet{Christopoulou:2001}&         $24$  &          $4$  &            $353$  &
        $43$ &                     $4.9$ \\
\citet{Pasachoff:2009} &          $49.2$  &         $40$  &            $426$  &                      $27$ &                     $7.2$ \\
This work (original)  &        $3.2-4.8$  &        $177$  &            $111$  &                      $60$ &                     $5.5$ \\
This work (degraded)  &           $38.4$  &         $52$\tablenotemark{c}  &   $340$  &          $26$ &                   $6.7$ \\
\enddata
\tablenotetext{a}{The two values refer to ``quiet'' and ``agitated'' regions, respectively.}
\tablenotetext{b}{The two values refer to equator and latitude $\pm 84^{\mathrm{o}}$, respectively.}
\tablenotetext{c}{Velocities were derived from 23 spicules.}
\tablecomments{Symbols denote: number of spicules $N$, lifetime $t$, apparent ascending velocity $v_{\mathrm{up}}$, and maximum height $H_{\mathrm{max}}$.}
\end{deluxetable}

\section{Discussion}

\subsection{Degraded Data and Classical Spicules}

Measuring spicules from degraded \emph{Hinode} data has a dramatic effect on their properties, as evidenced by Fig.~\ref{fig:hists}. Even more striking, the values from the degraded data agree very well with the classical results for spicules. It can be argued that our analysis suffers from systematic effects and uncertainties (like all studies of spicules, but which we minimize here as much as possible). However, the differential analysis between original and degraded data cancels those systematics and undeniably proves the effects of data degradation on spicule properties. Our results are not meant to provide the universal values of spicules measured from degraded data -- the rather approximate nature of our degradation precludes it -- but instead illustrate how such degradation \emph{can} affect measurements of spicules.

In Table~\ref{tab:spic_others} we compare our results with those of several ground-based studies, such as \citet{Roberts:1945}, \citet{Dizer:1952}, \citet{Rush:1954}, and \citet{Lippincott:1957}, which provided the basis for the reviews of \citet{Beckers:1968,Beckers:1972}. {Additionally, we include the results of \citet{Nishikawa:1988} and \citet{Christopoulou:2001}.
The more recent work of \citet{Pasachoff:2009} is also included, even though the authors used a modern telescope \citep[SST,][]{Scharmer:2003} with higher spatial resolution than \emph{Hinode}/SOT. Nevertheless, the authors used a very long cadence of $49.2\;$s and despite the high spatial resolution the observations were still subject to seeing effects in the time series.

All of the ground-based observations we compare with were made with H$\alpha$ filters, while the \emph{Hinode} data were observed in the \caii\ H filter. We are making the assumption that limb spicules should be comparable with both diagnostics. Spectroscopic studies show that H$\alpha$ and \caii\ H and K have similar Doppler velocities \citep{Athay:1964,Alissandrakis:1973}. Imaging studies such as the early work of \citet{Athay:1955} and most recently \citet{McIntosh:2008} show similar limb structures in H$\alpha$ and \caii\ filters. We used the same \emph{Hinode} SOT/NFI data as \citet{McIntosh:2008} to construct H$\alpha$ filtergrams and compare them with simultaneous SOT/BFI \caii\ H filtergrams. In Fig.~\ref{fig:halpha} we compare both filtergrams. These observations were taken starting at 14:46 UT on 2007 September 27. The H$\alpha$ filtergrams shown were obtained by summing NFI filtergrams at  H$\alpha \pm 80$~pm (red and blue wings, observed 3.2~s apart). It is clear from Fig.~\ref{fig:halpha} and the animation that both the spicule structure and the time evolution are very similar in H$\alpha$ and \caii\ H.
Thus, it seems reasonable to compare measurements from both diagnostics.

\begin{figure}
\begin{center}
\includegraphics[scale=0.95]{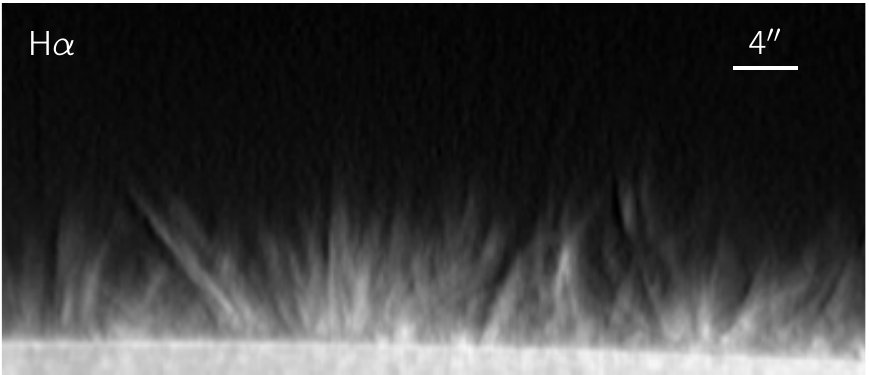}\\
\includegraphics[scale=0.95]{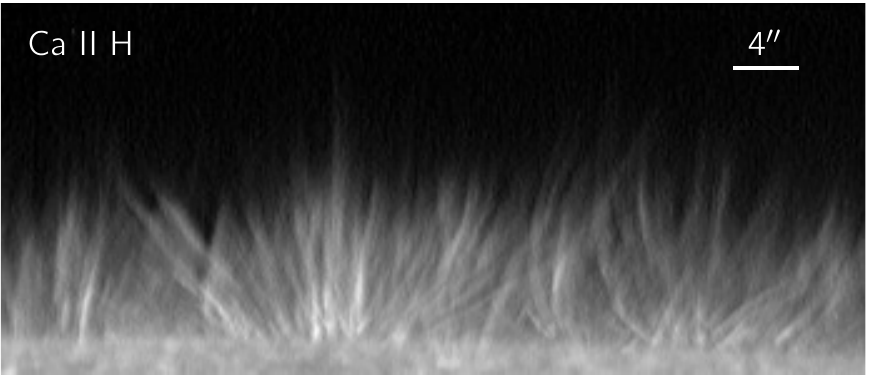}
\end{center}
\caption{H$\alpha$ and \caii\ H filtergrams from SOT polar observations taken at 14:46:32 UT on 2007 September 27. H$\alpha$ filtergram is the sum of red and blue wing NFI filtergrams (H$\alpha \pm 80$~pm, taken 3.2~s apart). Each \caii\ H BFI filtergram was observed 1.6~s after the second H$\alpha$ exposure. Radial density and emboss filters were applied to the images. Each panel is approximately $56\arcsec\times24\arcsec$, and the approximate solar (x, y) coordinates for the lower left corner are ($-2\arcsec.4$, $951\arcsec.9$). [\emph{This figure is available as an animation in the electronic edition of the Journal.}]\label{fig:halpha}}
\end{figure}

Many older studies do not distinguish between quiet sun, coronal holes, or active regions. With observations spanning a large area of the sun, the spicules studied were a sample of many regions. For our comparison we focused on quiet sun observations because the properties of quiet sun and coronal hole spicules are very similar \citepalias{Pereira:2012spic} and in the degraded data the difference would be negligible. Active regions were not included because they are not as common and were not singled out in the description of classical spicules.

With typical cadences between $30-60\;$s, older ground-based observations could only observe type II spicules in a few frames. It is perhaps not surprising that spicules with typical lifetimes of $50-150\;$s \citepalias{Pereira:2012spic} were generally not found. With the exception of \citet{Dizer:1952}, all of the listed studies give mean lifetimes above $200\;$s. In terms of mean ascending velocities, the first four studies in Table~\ref{tab:spic_others} have established the Beckers canonical value of $25\;\kms$. At the time, this value was accepted as representative also because it was consistent with the Doppler velocities derived from spectroscopic observations \citep{Beckers:1968}.
\citet{Nishikawa:1988} and \citet{Christopoulou:2001} find higher velocities, but from only 4 spicules each, all of which had parabolic trajectories (suggesting type I and not type II spicules).

Using higher spatial resolution observations, \citet{Pasachoff:2009} obtain results that agree very well with the classical values. While enormous progress has been made in  ground-based observations on the solar disk with adaptive optics and image reconstruction techniques \citep{vanNoort:2005}, the effects of seeing are harder to correct in limb images because of the lack of structures from which to derive the wavefront distortions. Given the long cadence used and the difficulties in correcting for seeing at the limb, it is perhaps not surprising that the measurements of \citet{Pasachoff:2009} are consistent with classical spicules.

\citet{Zhang:2012} claim that type II spicules may not even exist. Using \emph{Hinode} \caii\ H filtergrams the authors measure spicule properties and find most to be similar to type I spicules and thus compatible with the classical spicules. We disagree with their conclusions. In \citetalias{Pereira:2012spic} we analyze the same data sets and find a dominance of type II spicules, and inconsistencies within their measurements. With the present work we show that the existence of type II spicules is compatible with the measurements of classical spicules, given the lower resolution data used in the past.

After their rise phase, classical spicules have been reported to either fade or descend. The fraction of the spicules that descend (instead of fading) is not always reported, but can be more than half (\emph{e.g.} \citealt{Lippincott:1957} finds descent in  60\% of spicules, \citealt{Pasachoff:2009} only 28\%). In most of these cases the descent motions were irregular; \citet{Beckers:1968} notes that several studies ``notice the descent to be much more poorly defined than the ascent''. \citet{Rush:1954} note that the descent was often ``abrupt and irregular, suggesting that an apparent descent was actually a progressive loss of visibility''. This hints that at least some of the descents of spicules could be artifacts caused by seeing and low resolution; a ``loss of visibility'' is more consistent with the fading associated with type II spicules. In our original data analysis on \citetalias{Pereira:2012spic}, we find only 2\% of descending spicules in these data. In the degraded data we find no descending spicules, possibly owing to the lower number of detected spicules. This is at odds with the earlier reports. The reasons for this discrepancy are unclear. Some of the apparently descending spicules observed from the ground could be an artifact of seeing and resolution (as suggested by \citealt{Rush:1954}). It is also possible that these descents are less visible in \caii\ H than in H$\alpha$ because of the different passbands.

Both \citet{Nishikawa:1988} and \citet{Christopoulou:2001} find up and down parabolic motions in all 4 spicules measured, which suggests they are type I spicules. At least for the higher-quality observations of \citet{Christopoulou:2001}, it seems unlikely that the detection of parabolic motion results from data degradation. While \citetalias{Pereira:2012spic} finds a scarcity of type I spicules in quiet sun and coronal hole regions, it is unclear if there are nearby active regions in the observations of \citet{Nishikawa:1988} and \citet{Christopoulou:2001}. \citet{Christopoulou:2001} find a connection between observed spicules and mottles, further suggesting that the spicules studied are of type I, whose properties are consistent with classical spicules (see the discussion on type I spicules and mottles in \citetalias{Pereira:2012spic}).

\subsection{How Does Data Degradation Affect Spicule Measurements?}
Having established how the data degradation affects the measurements, it is important to investigate why. Looking at the degraded data and following the detected spicules, we identify several trends that affect the measurements. The clearest of those is what makes lifetimes longer. With a lower cadence and spatial resolution, following spicules is made difficult by their dynamic nature. It often happens that spicules recur from similar footpoints, clustered in a few regions, and with a high degree of superposition \citepalias{Pereira:2012spic}. Taking their transverse motions into account, when observed with long cadences it is often difficult to identify the same spicules in the next image. Adding a lower spatial resolution and many of these nearby spicules merge into each other, becoming extremely difficult to disentangle. Thus, the automated detection (and human eye) is fooled into thinking that a longer event is the same spicule, when in reality a much more dynamic scenario of spicules coming and going, and moving transversely occurs. Because now these spicules are measured over longer lifetimes, the likelihood of reaching larger heights increases because by bundling several spicules together we will measure the tallest in that group.
This shifts the height distribution to larger values (but does not affect the largest values of the distribution, as should be expected). The effect on the apparent velocities is more convoluted and not so clear cut. A possible explanation is that with longer cadences one tends to miss the initial rise phase of the spicules, which represents a smaller fraction of their lifetime. Hence, one would preferentially sample a smaller range of spicule heights that together with the longer lifetimes would yield a lower apparent velocity. In such conditions one should also expect to get several irregular space-time diagrams for the spicule lengths, where no clear ascent or descent of the spicule is visible. This is corroborated by the fact that we could not get a reliable measurement of velocity from about half of the detected spicules, because of such irregular diagrams. This was often also the case in classical studies, where several authors only derived velocities from a small fraction of the spicules used for height measurements \citep{Rush:1954,Lippincott:1957,Nishikawa:1988}. 

It is likely that lifetimes derived from classical studies of spicules are instead related to other dynamic timescales, such as spicule recurrence. It has been noted that spicules recur from the same footpoints (\citealt{Beckers:1968}; \citetalias{Pereira:2012spic}). Using transition region and coronal lines, \citet{McIntosh:2009} analyzed upflows believed to be associated with type II spicules and find that these upflows recur with a timescale around $3-15\;$min. This range is consistent with the lifetimes derived from classical spicules, suggesting that those lifetimes reflect not the lifetime of an individual spicule but the lifetime of a larger ensemble: the timescale of spicule recurrence. \citet{Sekse:2012b} find similar recurrence of Rapid Blueshifted Events, the likely disk-counterpart of type II spicules.

In this view, the classical features are actually conglomerates of (``type II'') spicules. This is compatible with the speculation by \citet{Zaqarashvili:2007} of ``micro-spicules'' constituting the coarser spicules detected with their high cadence and low resolution coronagraphic observations.

\section{Conclusions}

The recent discovery of two types of spicules \citep{DePontieu:2007} shows that spicules are more dynamic than previously thought. Type II spicules in particular have short lifetimes ($50-150\;$s), high ascending velocities ($30-130\;\kms$), and are dominant in most of the solar limb, except in active regions \citepalias{Pereira:2012spic}.

The properties of type II spicules (\citealt{DePontieu:2007}; \citetalias{Pereira:2012spic}) are incompatible with the classical views on spicules, consisting of lifetimes around 5 min and apparent velocities of $25\;\kms$ \citep{Beckers:1968,Beckers:1972}. Because type II spicules are likely to dominate in most locations where classical spicules were observed, it becomes a problem to reconcile both views. We suggest that these differences arise from a mix of spicule superposition and the low spatial and temporal resolution of older observations. To test this hypothesis we degrade \emph{Hinode}/SOT observations to similar resolution and cadence of older studies, and analyze the spicules using the same method as \citetalias{Pereira:2012spic}. Comparing the results of the original against the degraded data, we find that the image degradation has a significant effect on the measured spicule properties. For the degraded data we derive a mean lifetime around 5~min and a mean ascending velocity of $26\;\kms$, in very good agreement with the classical values. 

We conclude that type II spicules were likely previously detected in the studies reviewed by Beckers, but that their low spatial and temporal resolutions did not allow for an accurate determination of their dynamic properties.  This resolves the puzzling discrepancy between spicules seen from \emph{Hinode} and older ground-based studies, and firmly establishes that the classical properties are not representative. Theoretical studies of spicule formation, which so far have relied mostly on the classical spicule description, cannot afford to ignore the fundamentally different properties of type II spicules. These must be taken into account to provide model constraints.

\acknowledgements

TMDP was supported by the NASA Postdoctoral Program at Ames Research Center (NNH06CC03B). BDP was supported by NASA (NNX08AH45G, NNX08BA99G, and NNX11AN98G). This research was supported by the Research Council of Norway. The research leading to these results has received funding from the European Research Council under the European Union's Seventh Framework Programme (FP7/2007-2013) / ERC Grant agreement 291058. \emph{Hinode} is a Japanese mission developed by ISAS/JAXA, with the NAOJ as domestic partner and NASA and STFC (UK) as international partners. It is operated in cooperation with ESA and NSC (Norway). We thank Silje Bj\o lseth and Anne Fox for help with data reduction.

\bibliographystyle{apj}

\begin{thebibliography}{33}
\expandafter\ifx\csname natexlab\endcsname\relax\def\natexlab#1{#1}\fi

\bibitem[{{Alissandrakis}(1973)}]{Alissandrakis:1973}
{Alissandrakis}, C.~E. 1973, \solphys, 32, 345

\bibitem[{{Athay} \& {Bessey}(1964)}]{Athay:1964}
{Athay}, R.~G., \& {Bessey}, R.~J. 1964, \apj, 140, 1174

\bibitem[{{Athay} \& {Holzer}(1982)}]{Athay:1982}
{Athay}, R.~G., \& {Holzer}, T.~E. 1982, \apj, 255, 743

\bibitem[{{Athay} \& {Roberts}(1955)}]{Athay:1955}
{Athay}, R.~G., \& {Roberts}, W.~O. 1955, \apj, 121, 231

\bibitem[{{Beckers}(1968)}]{Beckers:1968}
{Beckers}, J.~M. 1968, \solphys, 3, 367

\bibitem[{{Beckers}(1972)}]{Beckers:1972}
---. 1972, \araa, 10, 73

\bibitem[{{Christopoulou} {et~al.}(2001){Christopoulou}, {Georgakilas}, \&
  {Koutchmy}}]{Christopoulou:2001}
{Christopoulou}, E.~B., {Georgakilas}, A.~A., \& {Koutchmy}, S. 2001, \solphys,
  199, 61

\bibitem[{{De Pontieu} {et~al.}(2009){De Pontieu}, {McIntosh}, {Hansteen}, \&
  {Schrijver}}]{DePontieu:2009}
{De Pontieu}, B., {McIntosh}, S.~W., {Hansteen}, V.~H., \& {Schrijver}, C.~J.
  2009, \apjl, 701, L1

\bibitem[{{De Pontieu} {et~al.}(2007){De Pontieu}, {McIntosh}, {Hansteen},
  {Carlsson}, {Schrijver}, {Tarbell}, {Title}, {Shine}, {Suematsu}, {Tsuneta},
  {Katsukawa}, {Ichimoto}, {Shimizu}, \& {Nagata}}]{DePontieu:2007}
{De Pontieu}, B., {McIntosh}, S., {Hansteen}, V.~H., {et~al.} 2007, \pasj, 59,
  655

\bibitem[{{De Pontieu} {et~al.}(2011){De Pontieu}, {McIntosh}, {Carlsson},
  {Hansteen}, {Tarbell}, {Boerner}, {Martinez-Sykora}, {Schrijver}, \&
  {Title}}]{DePontieu:2011}
{De Pontieu}, B., {McIntosh}, S.~W., {Carlsson}, M., {et~al.} 2011, Science,
  331, 55

\bibitem[{{Dizer}(1952)}]{Dizer:1952}
{Dizer}, M. 1952, Compt. Rend. Acad. Sci., 235, 1016

\bibitem[{{Kosugi} {et~al.}(2007){Kosugi}, {Matsuzaki}, {Sakao}, {Shimizu},
  {Sone}, {Tachikawa}, {Hashimoto}, {Minesugi}, {Ohnishi}, {Yamada}, {Tsuneta},
  {Hara}, {Ichimoto}, {Suematsu}, {Shimojo}, {Watanabe}, {Shimada}, {Davis},
  {Hill}, {Owens}, {Title}, {Culhane}, {Harra}, {Doschek}, \&
  {Golub}}]{Kosugi:2007}
{Kosugi}, T., {Matsuzaki}, K., {Sakao}, T., {et~al.} 2007, \solphys, 243, 3

\bibitem[{{Lippincott}(1957)}]{Lippincott:1957}
{Lippincott}, S.~L. 1957, Smithsonian Contributions to Astrophysics, 2, 15

\bibitem[{{McIntosh} \& {De Pontieu}(2009)}]{McIntosh:2009}
{McIntosh}, S.~W., \& {De Pontieu}, B. 2009, \apj, 707, 524

\bibitem[{{McIntosh} {et~al.}(2008){McIntosh}, {De Pontieu}, \&
  {Tarbell}}]{McIntosh:2008}
{McIntosh}, S.~W., {De Pontieu}, B., \& {Tarbell}, T.~D. 2008, \apjl, 673, L219

\bibitem[{{Nishikawa}(1988)}]{Nishikawa:1988}
{Nishikawa}, T. 1988, \pasj, 40, 613

\bibitem[{{Okamoto} {et~al.}(2007){Okamoto}, {Tsuneta}, {Berger}, {Ichimoto},
  {Katsukawa}, {Lites}, {Nagata}, {Shibata}, {Shimizu}, {Shine}, {Suematsu},
  {Tarbell}, \& {Title}}]{Okamoto:2007}
{Okamoto}, T.~J., {Tsuneta}, S., {Berger}, T.~E., {et~al.} 2007, Science, 318,
  1577

\bibitem[{{Pasachoff} {et~al.}(2009){Pasachoff}, {Jacobson}, \&
  {Sterling}}]{Pasachoff:2009}
{Pasachoff}, J.~M., {Jacobson}, W.~A., \& {Sterling}, A.~C. 2009, \solphys,
  260, 59

\bibitem[{{Pereira} {et~al.}(2012){Pereira}, {De Pontieu}, \&
  {Carlsson}}]{Pereira:2012spic}
{Pereira}, T.~M.~D., {De Pontieu}, B., \& {Carlsson}, M. 2012, \apj, 759, 18

\bibitem[{{Pneuman} \& {Kopp}(1978)}]{PneumanKopp:1978}
{Pneuman}, G.~W., \& {Kopp}, R.~A. 1978, \solphys, 57, 49

\bibitem[{{Roberts}(1945)}]{Roberts:1945}
{Roberts}, W.~O. 1945, \apj, 101, 136

\bibitem[{{Rouppe van der Voort} {et~al.}(2009){Rouppe van der Voort},
  {Leenaarts}, {de Pontieu}, {Carlsson}, \& {Vissers}}]{vdVoort:2009}
{Rouppe van der Voort}, L., {Leenaarts}, J., {de Pontieu}, B., {Carlsson}, M.,
  \& {Vissers}, G. 2009, \apj, 705, 272

\bibitem[{{Rush} \& {Roberts}(1954)}]{Rush:1954}
{Rush}, J.~H., \& {Roberts}, W.~O. 1954, Australian Journal of Physics, 7, 230

\bibitem[{{Scharmer} {et~al.}(2003){Scharmer}, {Bjelksjo}, {Korhonen},
  {Lindberg}, \& {Petterson}}]{Scharmer:2003}
{Scharmer}, G.~B., {Bjelksjo}, K., {Korhonen}, T.~K., {Lindberg}, B., \&
  {Petterson}, B. 2003, Proc. SPIE, 4853, 341

\bibitem[{{Secchi}(1877)}]{secchi1877}
{Secchi}, A. 1877, Le Soleil, Vol.~2 (Paris: Gauthier-Villars)

\bibitem[{{Sekse} {et~al.}(2012){Sekse}, {Rouppe van der Voort}, \& {De
  Pontieu}}]{Sekse:2012b}
{Sekse}, D.~H., {Rouppe van der Voort}, L., \& {De Pontieu}, B. 2012, \apj,
  752, 108

\bibitem[{{Sterling}(2000)}]{Sterling:2000}
{Sterling}, A.~C. 2000, \solphys, 196, 79

\bibitem[{{Suematsu} {et~al.}(2008){Suematsu}, {Tsuneta}, {Ichimoto},
  {Shimizu}, {Otsubo}, {Katsukawa}, {Nakagiri}, {Noguchi}, {Tamura}, {Kato},
  {Hara}, {Kubo}, {Mikami}, {Saito}, {Matsushita}, {Kawaguchi}, {Nakaoji},
  {Nagae}, {Shimada}, {Takeyama}, \& {Yamamuro}}]{Suematsu:2008}
{Suematsu}, Y., {Tsuneta}, S., {Ichimoto}, K., {et~al.} 2008, \solphys, 249,
  197

\bibitem[{{Tsiropoula} \& {Tziotziou}(2004)}]{Tsiropoula:2004}
{Tsiropoula}, G., \& {Tziotziou}, K. 2004, \aap, 424, 279

\bibitem[{{Tsuneta} {et~al.}(2008){Tsuneta}, {Ichimoto}, {Katsukawa}, {Nagata},
  {Otsubo}, {Shimizu}, {Suematsu}, {Nakagiri}, {Noguchi}, {Tarbell}, {Title},
  {Shine}, {Rosenberg}, {Hoffmann}, {Jurcevich}, {Kushner}, {Levay}, {Lites},
  {Elmore}, {Matsushita}, {Kawaguchi}, {Saito}, {Mikami}, {Hill}, \&
  {Owens}}]{Tsuneta:2008}
{Tsuneta}, S., {Ichimoto}, K., {Katsukawa}, Y., {et~al.} 2008, \solphys, 249,
  167

\bibitem[{{van Noort} {et~al.}(2005){van Noort}, {Rouppe van der Voort}, \&
  {L{\"o}fdahl}}]{vanNoort:2005}
{van Noort}, M., {Rouppe van der Voort}, L., \& {L{\"o}fdahl}, M.~G. 2005,
  \solphys, 228, 191

\bibitem[{{Zaqarashvili} {et~al.}(2007){Zaqarashvili}, {Khutsishvili},
  {Kukhianidze}, \& {Ramishvili}}]{Zaqarashvili:2007}
{Zaqarashvili}, T.~V., {Khutsishvili}, E., {Kukhianidze}, V., \& {Ramishvili},
  G. 2007, \aap, 474, 627

\bibitem[{{Zhang} {et~al.}(2012){Zhang}, {Shibata}, {Wang}, {Mao}, {Matsumoto},
  {Liu}, \& {Su}}]{Zhang:2012}
{Zhang}, Y.~Z., {Shibata}, K., {Wang}, J.~X., {et~al.} 2012, \apj, 750, 16

\end{thebibliography}

\end{document}